\documentclass[final]{article}
\usepackage{lineno,hyperref}
\usepackage[dvipsnames]{xcolor}

\usepackage{amsmath}
\usepackage{amssymb}
\usepackage{mathtools}
\usepackage{graphicx,subfigure}
\usepackage{color}
\usepackage{enumerate}
\usepackage{caption}
\usepackage{ulem}

\renewcommand\vec[1]{\mathbf{#1}}

\begin{document}

\title{A general mathematical framework for understanding the behavior of heterogeneous stem cell regeneration}
\author{Jinzhi Lei\\ \\ Zhou Pei-Yuan Center for Applied Mathematics\\ Tsinghua University, Beijing 100084, China}

\maketitle

\begin{abstract}
 
Stem cell heterogeneity is essential for homeostasis in tissue development. This paper establishes a general mathematical framework to model the dynamics of stem cell regeneration with cell heterogeneity and random transitions of epigenetic states. The framework generalizes the classical G0 cell cycle model and incorporates the epigenetic states of individual cells represented by a continuous multidimensional variable. In the model,  the kinetic rates of cell behaviors, including proliferation, differentiation, and apoptosis, are dependent on their epigenetic states, and the random transitions of epigenetic states between cell cycles are represented by an inheritance probability function that describes the conditional probability of cell state changes. Moreover, the model can be extended to include genotypic changes and describe the process of gene mutation-induced tumor development. The proposed mathematical framework provides a generalized formula that helps us to understand various dynamic processes of stem cell regeneration, including tissue development, degeneration, and abnormal growth.
\end{abstract}



\section{Introduction}

Stem cell regeneration is an essential biological process in most self-renewing tissues during development and the maintenance of tissue homeostasis. Stem cells multiply by cell division, during which DNA is replicated and assigned to the two daughter cells along with the inheritance of epigenetic information and the partition of molecules. Unlike the accumulated process of DNA replication, the inheritance of epigenetic information is often subjected to random perturbations; for example, the reconstruction of histone modifications and DNA methylations is an intrinsically random processes of writing and erasing the modified markers \cite{Probst:2009iq, Wu:2014gw}, and the stochastic gene expression that yields phenotypic cell-to-cell variability within clonal populations \cite{Chang:2008gua,Wada:2018bo}. The stochastic inheritance of epigenetic changes during cell division can lead to different aspects of stem cell heterogeneity that are important for the dynamic equilibrium of various phenotypic cells during tissue development. The accumulation of undesirable epigenetic changes may result in abnormal growth and various types of serious diseases \cite{Farlik:2016ig, Field:2018io, Horvath:2013baa, Kulis:2010hz, McGranahan:2017jb, Meissner:2010ica, Portela:2010fg, Rudenko:2014ir, 2013Sci...339.1567S, Xie:2018hm}.

The heterogeneity of stem cells has been highlighted in recent years due to new technologies with single-cell resolution, which have led to the discovery of new cell types and changes in the understanding of differentiation landscapes \cite{Buettner:2015hp, Butler:2018ex, Haber:2017dt, Levitin:2018jb, Li:2017iv, Plass:2018in}. During early embryonic development, heterogeneous expression and histone modifications are correlated with cell fate and the dynamic equilibrium of pluripotent stem cells \cite{Hawkins:2011ht, Hayashi:2008fu, Parfitt:2010fh, TorresPadilla:2007jr}. Chromatin modifications in the human primary hematopoietic stem cell/progenitor cell (HSC/HPC) stage can lead to the dynamic equilibrium of heterogeneous and interconvertible HSCs \cite{Chang:2008gua, 2014NatSR...4E5199W}, as well as gene expression changes during differentiation \cite{Cui:2009fta}. Moreover, applications of single-cell RNA sequencing have revealed the continuous spectrum of differentiation in zebrafish \cite{Macaulay:2016bz}, mice \cite{Nestorowa:2016iq}, and human HSCs \cite{Velten:2017jj}. These findings have challenged the demarcation between stem cells and progenitor cells and have led to an evolving understanding of the complex hematopoietic differentiation landscape \cite{2018Natur.553..418L, Notta:2016hj}.

Heterogeneity plays an important role in the development of drug resistance. Cancer development is driven by the evolutionary selection of somatic genetic alterations and epigenetic alterations, which result in multistage tumorigenesis and heterogeneous cancer cell phenotypes \cite{Gerlinger:2012jt, Giustacchini:2017ic, JamalHanjani:2017cl, Li:2017iv,  Lim:2017fw, 2011Natur.472...90N, Wu:2017ed}. Tumors with different subtypes often differ in terms of treatment responses and patient survival \cite{DeSousaEMelo:2013hx, Giustacchini:2017ic, Sadanandam:2013fv}, and treatment stresses may also induce cancer cell plasticity and drug resistance \cite{Graf:2002tj, LeMagnen:2018fx, Marjanovic:2012fx, Soundararajan:2018bt, Tam:2013ip}. Cell plasticity is often associated with epigenetic modifications, and targeting the epigenetic regulators, such as the polycomb group protein EZH2, has been an attractive strategy in cancer treatment \cite{Fraietta:2018ira, Simon:2008iw, Topper:2017cd}. To better understand the progress of tumorigenesis and drug resistance, we need to develop predictive models for the evolutionary dynamics of cancer, including the regeneration of cancer stem cells \cite{Beerenwinkel:2016iu, Hanahan:2000hx,Hamis:2019eu}.

Despite the central role of stem cell regeneration in tissue development, quantitative investigation of the process is well beyond the ability of current technologies. Furthermore, in many fields of biological science, mathematical modeling tools have aided in improving the understanding of the principles of related processes \cite{Altrock:2015dt, Lander:2013hb, 2012Sci...336..187M, Morris:2014cp}. In 2007, Weinberg posed the following question \cite{2007Natur.449..978W}: ``can algebraic formulae tell us more than reasoning about the behavior of complex biological systems?'' Various computational models have been developed in studies of tissue development and cancer systems biology under different circumstances and hypotheses \cite{Beerenwinkel:2016iu, Beerenwinkel:2015eb, Du:2018gb, Greene:2016du, Greulich:2016ju, Guo:2017ix, Werner:2011dk, Werner:2016gi}. Nevertheless, a unified formulation that bypasses detailed assumptions is required to provide a more basic logic of the biological behaviors of these complex processes. In this paper, based on the general process of the cell cycle and heterogeneous stem cell regeneration, we propose a general mathematical framework to formulate the dynamics of heterogeneous stem cell regeneration. The model framework includes essential cellular behaviors, including proliferation, apoptosis, and differentiation/senescence; however, it bypasses the biological details of signaling pathways but emphasizes stem cell heterogeneity and epigenetic inheritance during cell cycling. Further extensions of the model framework with different formulas for the kinetic rates and the inheritance probability can be applied to various biological processes, such as embryonic development, tissue disease and degeneration, and tumor development.

The aim of this paper was to introduce a new general formula for the dynamics of stem cell regeneration with an emphasis on the effects of cell heterogeneity; therefore, a discussion of concrete conclusions based on the formula was not included. The simulation results below were included to demonstrate the potential applications of the model and were not related to any actual biological processes. 

\section{Results}
\subsection{The G0 cell cycle model for homogeneous stem cell regeneration}

\begin{figure}[htbp]
\centering
\includegraphics[width=10cm]{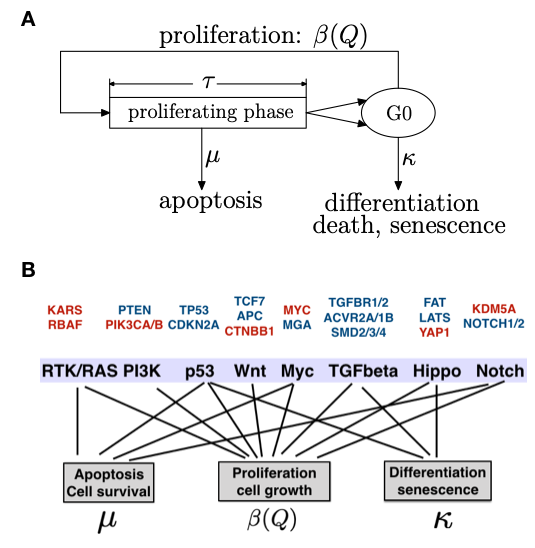}
\caption{\textbf{G0 cell cycle model for homogeneous stem cell regeneration.} \textbf{(A)}. A schematic of the G0 model of stem cell regeneration. During stem cell regeneration, cells in the resting phase either enter the proliferation phase with rate $\beta$ or are removed from the resting pool with rate $\kappa$ due to differentiation, aging, or death. Proliferating cells undergo apoptosis with rate $\mu$. \textbf{(B)}. Oncogenic signaling pathways and their associated cell behaviors and parameters in the G0 model. For each pathway, the genes are highly altered (according to the dataset in the TCGA PanCancer Atlas) by oncogenic activator (red) and tumor suppressor inactivators (blue). Details of the pathways and genes are described by \cite{SanchezVega:2018jw}.}
\label{fig:1}
\end{figure}

A classical model used to describe the dynamics of stem cell regeneration is the G0 cell cycle model proposed in the 1970s \cite{Burns:1970tm, Mackey:1978vy}. In this model, homogeneous cell cycles are classified into resting (G0) or proliferating (G1, S, and G2 phases and mitosis) phases (Figure \ref{fig:1}A). During each cell cycle, a cell in the proliferating phase either undergoes apoptosis or divides into two daughter cells; however, a cell in the resting phase either irreversibly differentiates into a terminally differentiated cell or returns to the proliferating phase. This process can be modeled by an age-structure model for cell numbers in the resting phase and the proliferating phase. Integrating the age-structure model through the characteristic line method gives the following delay differential equation (Material and methods)
\begin{equation}
\label{eq:1}
\dfrac{d Q}{d t} = - (\beta(Q) + \kappa) Q + 2 e^{-\mu \tau} \beta(Q_\tau) Q_\tau.
\end{equation}
Here,  $Q$ is the cell counts at the resting phase, $\beta(Q)$ is the proliferation rate, $\mu$ is the apoptosis rate of the cells in the proliferating phase, $\tau$ is the duration of the proliferating phase, and $\kappa$ is the rate of removing cells out of the resting phase, which includes terminal differentiation, cell death, or senescence (hereafter, we call $\kappa$ the differentiation rate for simplicity). The subscript indicates a time delay, \textit{i.e.}, $Q_\tau$ indicates $Q(t-\tau)$. The proliferation rate $\beta(Q)$ describes how cells regulate the self-renewal of stem cells through secreted cytokines and is often given by a decreasing function, hence $\beta(0) = \beta_0 \leq \beta(Q) \leq \beta_\infty = \lim_{Q\to+\infty} \beta(Q)$ (Material and method). Typically, for normal individuals, we usually have $\beta_\infty = 0$ because of the inhibition of the cell cycle pathways.

The G0 cell cycle model and its extensions are widely used to investigate hematopoietic stem cell dynamics \cite{Adimy:2014gs, Foley:2009jz, Jinzhi:2011bl,  Mackey:2001ux}; dysregulation of the apoptosis rate or differentiation rate of hematopoietic stem cells can result in serious periodic hematopoietic diseases \cite{Dale:2015ia}. Moreover, from Eq. \eqref{eq:1}, the stem cell dynamics are mainly determined by the pathways related to stem cell proliferation, apoptosis, differentiation, senescence, and growth. Major oncogenic signaling pathways obtained from an integrated analysis of genetic alterations in The Cancer Genome Atlas (TCGA) \cite{SanchezVega:2018jw} show direct connections to the coefficients $\beta(Q),\mu$, and $\kappa$ in \eqref{eq:1} (Figure \ref{fig:1}B) (Material and methods). Eq. \eqref{eq:1} is capable of describing the population dynamics of stem cell regeneration. Nevertheless, cell heterogeneity is not included in the model, however, it has been highlighted in recent years for its importance in understanding cancer development and drug resistance in cancer therapy. 

\subsection{The general framework of heterogeneous stem cell regeneration}

\begin{figure}[htbp]
\centering
\includegraphics[width=10cm]{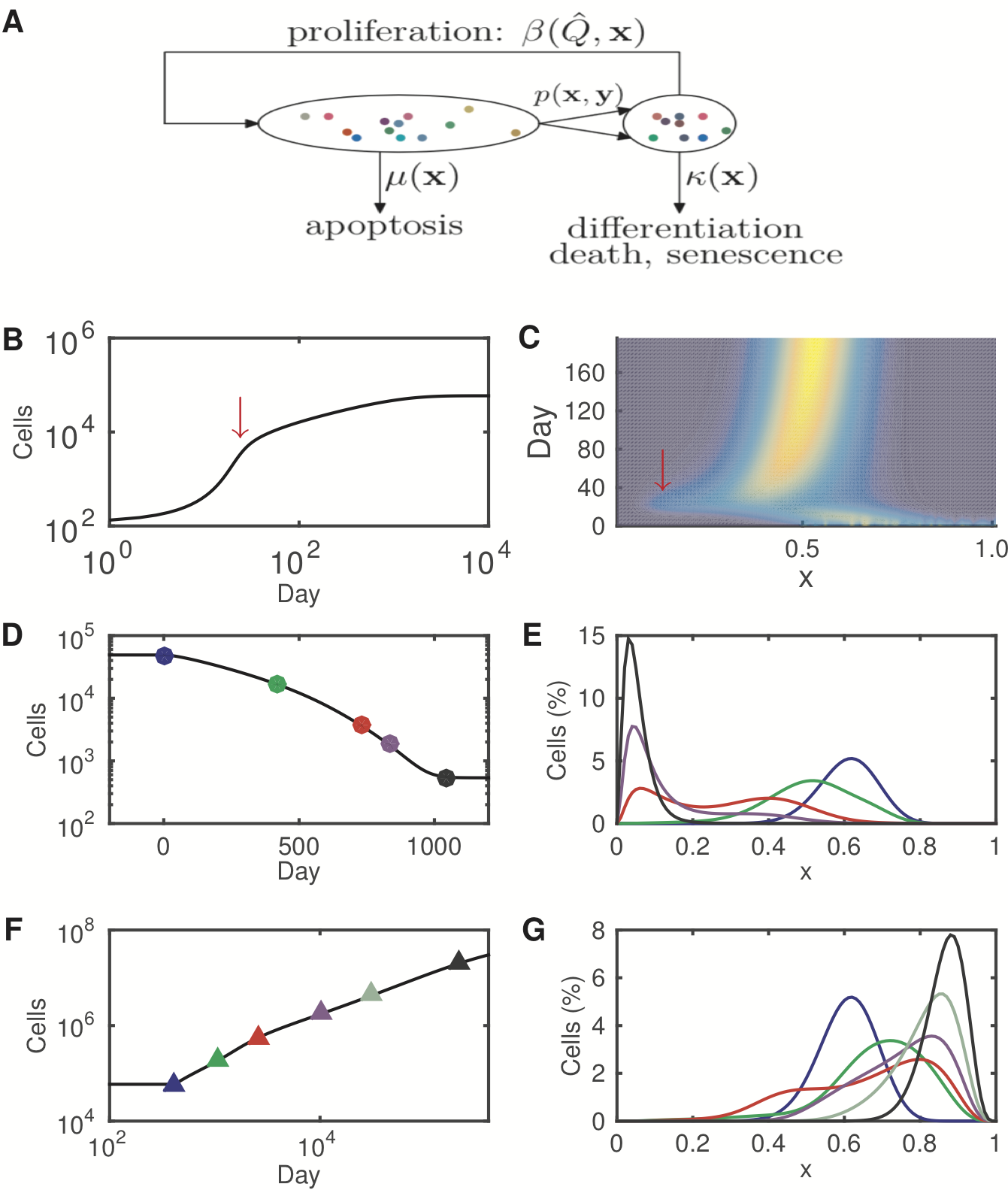}
\caption{\textbf{Heterogeneous stem cell regeneration.} \textbf{(A)}. A schematic diagram of the model of heterogeneous stem cell regeneration. Cell heterogeneity is represented by the epigenetic state ($\vec{x}$), indicated by dots with different colors. The dynamic rates for each cell are dependent on the epigenetic state, which varies during cell division according to the inheritance probability $p(\vec{x},\vec{y})$. \textbf{(B)}. Simulated population dynamics for the growth process with the increasing numbers of cells toward a steady state. \textbf{(C)}. The simulated evolution of the cell counts with varying heterogeneity in the growth process. The red arrows in (B) and (C) indicate a temporary increase in a subpopulation of cells with a low level of $x$ at the early stage. \textbf{(D)}. Simulated population dynamics of the degeneration process with decreasing numbers of cells. \textbf{(E)}. The percentage of cells with different epigenetic states during the degeneration process. Timepoints for each curve are indicated by dots with the same colors as those shown in (D). \textbf{(F)}. Simulated population dynamics of the abnormal growth process with increasing numbers of cells. \textbf{(G)}. The percentage of cells with different epigenetic states during the abnormal growth process. Timepoints for each curve are indicated by triangles with the same colors as those shown in (F). See Material and methods for the simulation details.}
\label{fig:2}
\end{figure}

To extend the abovementioned G0 cell cycle model to include cell heterogeneity, we introduced a quantity $\vec{x}$ (scalar or vector) that refers to the epigenetic state of a cell and denoted $Q(t,\vec{x})$ as the cell count at time $t$ with state $\vec{x}$ (Figure \ref{fig:2}A). In general, $\vec{x}$ can refer to any quantity associated with a cell, such as the expression levels of marker genes, histone modifications in nucleosomes, or DNA methylations. Experimentally, the quantities can be measured by single-cell sequencing techniques, such as single-cell RNA sequencing\cite{Tang:2009kj}, single-cell ChIP sequencing\cite{Rotem:2015bp}, or single-cell genome-wide bisulfite sequencing\cite{Smallwood:2014kn}. In applications, we usually do not use $\vec{x}$ to refer to the whole genome data. Instead, we may consider $\vec{x}$ as low dimensional quantities associated with specific genes that may affect the signaling pathways controlling cell cycle progression, apoptosis, or cell growth. Hence, the kinetic rates $\beta$, $\mu$, $\kappa$, and the duration of the proliferating phase $\tau$ in Eq. \eqref{eq:1} are cell specific and dependent on the state $\vec{x}$ of each individual cell. Moreover, all cells in the niche can interfere with stem cell self-renewal through the released cytokines. Let $\xi(\vec{x})$ denote the effective cytokine signal produced by a cell with state $\vec{x}$, and let $c=\int Q(t,\vec{x}) \xi(\vec{x}) d \vec{x}$ denote the total concentration of effective cytokines that regulate cell proliferation. The proliferation rate in Eq. \eqref{eq:1} may depend on the concentration $c$, and hence $\beta$ is rewritten as $\beta(c,\vec{x})$. 

When cell-to-cell variability is considered, the inheritance of the epigenetic states of cells during cell division is essential to shape the distribution of cell heterogeneity. Many biological processes, such as the random partition of molecules \cite{Huh:2011gc}, the random inheritance of nucleosome modification \cite{Dodd:2007em, Probst:2009iq} and DNA methylation \cite{Wu:2014gw, Horvath:2013baa}, are involved in shaping the inheritance of epigenetic states from mother to daughter cells after cell division. Many efforts have been made to model epigenetic cell memory \cite{Dodd:2007em, Haerter:2014er, Huang:2017jr, Huh:2011gc, 2017IJMPB..3150243S}; however, it remains challenging to develop precise models for this process, which is not yet clear. Nevertheless, while we can overlook the biological details and focus on the changes of epigenetic states, we can also introduce the inheritance probability $p(\vec{x},\vec{y})$, which represents the probability that a daughter cell of state $\vec{x}$ comes from a mother cell of state $\vec{y}$ after cell division. Therefore, $p(\vec{x}, \vec{y})$ describes the overall effect of epigenetic state inheritance after cell division, and $p(\vec{x},\vec{y}) d \vec{x} = 1$ for any $\vec{y}$. Based on the above assumptions and the similar argument in the homogeneous model, the dynamical equation for $Q(t,\vec{x})$ is given by (Material and methods):
\begin{equation}
\label{eq:2}
\left\{
\begin{array}{rcl}
\dfrac{\partial Q(t,\vec{x})}{\partial t} &=& - Q(t,\vec{x}) (\beta(c,\vec{x}) + \kappa(\vec{x}))\\
&&{}\displaystyle + 2 \int \beta(c_{\tau(\vec{y})}, \vec{y})Q(t-\tau(\vec{y}),\vec{y})e^{-\mu(\vec{y})\tau(\vec{y})}p(\vec{x},\vec{y})d\vec{y}\\
c(t) &=&\displaystyle \int Q(t,\vec{x}) \xi(\vec{x}) d \vec{x}.
\end{array}
\right.
\end{equation}
Here, the integrals are taken over all possible epigenetic states. Moreover, if we consider discrete states, such as gene mutations, we can extend the integrals to include the summation over all discrete states (to be detailed below). Eq. \eqref{eq:2} extends the previous G0 cell cycle model and provides a general framework for heterogeneous stem cell regeneration.

Eq. \eqref{eq:2} is an autonomous system in which the rate functions and inheritance probability are not explicitly dependent on time $t$. Nevertheless, while we apply the equation to situations with time dependence, such as embryo development, environmental changes, injury, and external stimuli, the time-dependent rate functions and the inheritance probability can be included in a straightforward manner. An example is shown in Figure \ref{fig:5} below.

\begin{figure}[thbp]
\centering
\includegraphics[width=6cm]{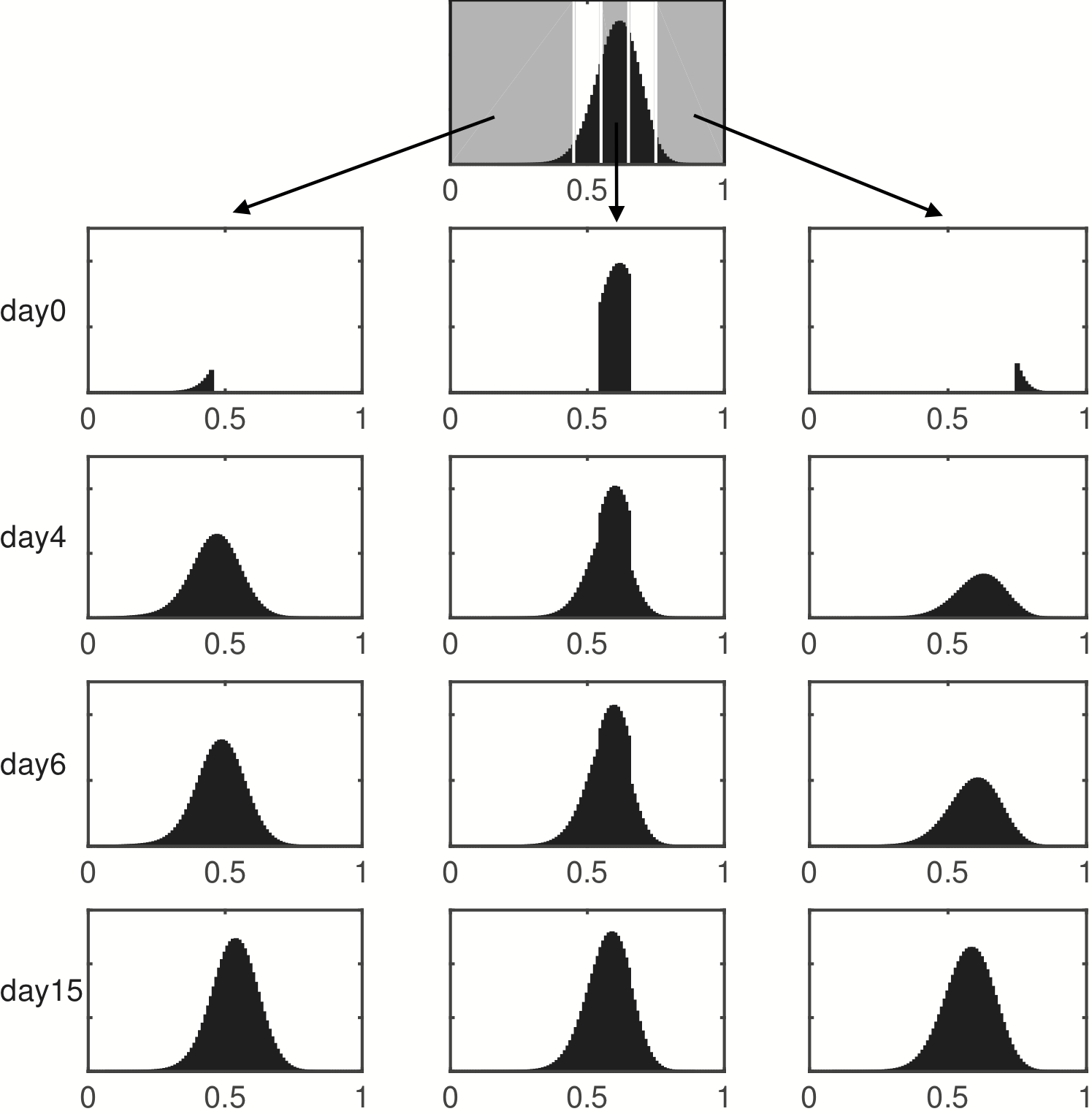}
\caption{\textbf{Restoration of heterogeneity from a fraction of the cell subpopulation.} Clonal cells with the highest ($x>0.75$), middle ($0.55<x<0.65$), and lowest ($x<0.45$) epigenetic states independently re-established the parental extent of clonal heterogeneity in separate simulations.}
\label{fig:3}
\end{figure}

Based on the Eq. \eqref{eq:2}, when we introduced appropriate definitions for the kinetic rates and the inheritance function, we are able to model various biological processes of stem cell regeneration, such as tissue growth, degeneration, and abnormal growth (Figure \ref{fig:2}B-G). For simplicity, we consider the situations with a single epigenetic state $x$ ($0\leq x \leq1$) that affects only cell proliferation and differentiation in a manner similar to the stemness so that a larger value of $x$ indicates higher stemness (Material and methods). Figure \ref{fig:2}B-C show the dynamics of tissue growth starting from a small population of cells with high levels of stemness. There is a temporary transition at the early stage characterized by a rapid increase in the cell number and a subpopulation of cells with a low level of stemness (Figure \ref{fig:2}B-C, red arrows). Figure \ref{fig:2}D-E show the dynamics of degeneration with alterations in the inheritance function, and Figure \ref{fig:2}F-G show the abnormal growth due to a decrease in the differentiation rate and an alteration in the inheritance function. The processes of both degeneration and abnormal growth include a short-term stage of biphenotypic  cell subpopulations with either high or low stemness (Figures \ref{fig:2}E and G, red curves). Moreover, simulations show that the steady state heterogeneity can be restored from cell subpopulation fractions (Figure \ref{fig:3}), which is in agreement with experiments that were previously explained by transcriptome-wide noise \cite{Hoffmann:2008cx, 2016PNAS..113.2672L, 2014NatSR...4E5199W}.

Eq. \eqref{eq:2} describes the evolution of the cell numbers with various epigenetic states, according to which the total cell number $Q(t) = \int Q(t, \vec{x})d\vec{x}$ and the density of cells with different epigenetic states $f(t, \vec{x}) = Q(t, \vec{x})/Q(t)$ are relevant to the experimental observations. From Eq. \eqref{eq:2}, it is easy to obtain the equations for $Q(t)$ and $f(t, \vec{x})$ as follows (see Materials and methods):
\begin{eqnarray}
\label{eq:3}
\dfrac{d Q}{d t} &=& - \int Q(t,\vec{x}) (\beta(c,\vec{x}) + \kappa(\vec{x})) d \vec{x} \\
&&{}+ 2 \int \beta(c_{\tau(\vec{x})},\vec{x}) Q(t-\tau(\vec{x}),\vec{x}) e^{-\mu(\vec{x})\tau(\vec{x})} d \vec{x}, \nonumber
\end{eqnarray}
and
\begin{eqnarray}
\label{eq:4}
\dfrac{\partial f(t,\vec{x})}{\partial t} &=& -f(t,\vec{x}) \int f(t,\vec{y}) ((\beta(c,\vec{x}) + \kappa(\vec{x}))- (\beta(c,\vec{y}) + \kappa(\vec{y})))d\vec{y}\\
&&{} + \dfrac{2}{Q(t)}\int \beta(c_{\tau(\vec{y})},\vec{y})Q(t-\tau(\vec{y}),\vec{y})e^{-\mu(\vec{y})\tau(\vec{y})}(p(\vec{x},\vec{y})-f(t,\vec{x}))d\vec{y}.\nonumber
\end{eqnarray}
Eqs. \eqref{eq:3} and \eqref{eq:4} provide the evolution dynamics of relative  numbers of cells that can be obtained from experiments by single-cell sequencing or flow cytometry. Here we note that when $\xi(\vec{x}) = 1$, we have $c(t)=Q(t)$, and hence, Eqs. \eqref{eq:3}-\eqref{eq:4} provide a closed-form equation.

Here, the state variable $\vec{x}$ represents the epigenetic state of an individual cell, and $p(\vec{x}, \vec{y})$ represents the inheritance function; hence, Eq. \eqref{eq:2} describes the dynamics with epigenetic state transitions. Nevertheless, the proposed framework can also be extended to describe the changes in genetic alternations if we consider $\vec{x}$ as the genetic state and $p(\vec{x},\vec{y})$ as the probability of point mutations. This is often the situation underlying the genomic instability associated with cancer development \cite{2013Natur.501..338B, Hanahan:2011gu, Sobinoff:2017gb, Zhivotovsky:2004ec}, and hence, the model can be used to study genetic heterogeneity in cancer development. In this paper, we mainly focus on the equation including epigenetic state transitions and assume that $\vec{x}$ always represents the epigenetic state. 

\subsection{Stochastic epigenetic state inheritance during cell divisions}

In Eqs. \eqref{eq:2}-\eqref{eq:4}, the mathematical formula for epigenetic state-relevant coefficients should be given based on how the epigenetic states (or genes) may affect the relative biological process. In the equation, the kinetic rates are associated with cell population dynamics; however, it is in general difficult to determine the inheritance function $p(\vec{x},\vec{y})$, which may depend on the biochemical details within cell divisions. Here, we proposed a phenomenological inheritance function to represent the stochastic inheritance of the epigenetic states. More specifically, let $\vec{x}=(x_1, x_2,\cdots,x_n)$ represent the expression levels of $n$ marker genes; we then derive the inheritance function $p_i(x_i,\vec{y})$ for each gene so that $p(\vec{x},\vec{y})=\prod_{i=1}^n p_i(x_i,\vec{y})$.

We assume that the epigenetic state of a daughter cell is a random number with a distribution depending on that of the mother cell. In previous studies based on stochastic simulations of gene expressions coupled with nucleosome modifications over multiple cell cycles \cite{Huang:2017jr, Jiao:2018gy}, we found that the nucleosome modification level of daughter cells, when normalized to the interval $[0,1]$, can be well described by a beta-distributed random number dependent on the state of the mother cell. Therefore, we generalize these findings and define the inheritance function $p_i(x_i,\vec{y})$ through the beta distribution density function as follows:
\begin{equation}
\label{eq:5-0}
p_i (x_i,\vec{y})=\dfrac{x_i^{a_i(\vec{y})-1} (1-x_i)^{b_i(\vec{y})-1}}{B(a_i(\vec{y}),b_i (\vec{y}))},\quad B(a,b)=\dfrac{\Gamma(a)\Gamma(b)}{\Gamma(a+b)},
\end{equation}
where $\Gamma(z)$ is the gamma function, and $a_i$ and $b_i$ are shape parameters that depend on the state of the mother cell. If we assume that the mean and variance of $x_i$, given the state $\vec{y}$, are
$$ 
\mathrm{E}(x_i | \vec{y})=\phi_i(\vec{y}),\ \mathrm{Var}(x_i | \vec{y})=\dfrac{1}{1+\eta_i(\vec{y})} \phi_i(\vec{y})(1 - \phi_i(\vec{y})),
$$
the shape parameters are given by (Material and methods)
\begin{equation}
\label{eq:5}
a_i(\vec{y}) = \eta_i(\vec{y})\phi_i(\vec{y}),\quad b_i (\vec{y}) = \eta_i(\vec{y})(1 - \phi_i (\vec{y})).
\end{equation}
Here, we note that $\phi_i(\vec{y})$ and $\eta_i(\vec{y})$ always satisfy
\begin{equation}
\label{eq:6}
0<\phi_i(\vec{y})<1, \   \eta_i(\vec{y})>0.
\end{equation}
Hence, the inheritance function can be determined through the predefined functions $\phi_i(\vec{y})$ and $\eta_i(\vec{y})$, often through data-driven modeling or assumptions, that satisfy Eq. \eqref{eq:6}.  

From the above argument, given the functions $\phi_i(\vec{y})$ and $\eta_i(\vec{y})$ that satisfying Eq. \eqref{eq:6}, and the shape parameters $a_i(\vec{y})$ and $b_i(\vec{y})$ defined by Eq. \eqref{eq:5}, the inheritance probability $p(\vec{x}, \vec{y})$ can be given by
\begin{equation}
\label{eq:p9}
p(\vec{x}, \vec{y}) = \prod_{i=1}^n\dfrac{x_i^{a_i(\vec{y})-1} (1- x_i)^{b_i(\vec{y})-1}}{B(a_i(\vec{y}), b_i(\vec{y}))}.
\end{equation}

Given the inheritance probability, we can numerically solve Eq. \eqref{eq:2} to obtain the dynamics of cell growth. However, when there are two or more epigenetic states, it is very expensive to solve the high dimensional integral equation \eqref{eq:2}. Hence, instead of solving the equation directly, we can perform a single-cell-based stochastic simulation based on the above mathematical model. In the stochastic simulation,  we model the random proliferation, apoptosis, and terminal differentiation of individual cells in a multiple cell system. In the system, each cell has its own epigenetic state and randomly undergoes proliferation, apoptosis, or terminal differentiation with a probability depending on the epigenetic state. When a cell undergoes mitosis, the cell divides into two cells, and the epigenetic states of the two daughter cells are calculated based on the above inheritance probability functions (see Material and methods).

\begin{figure}[htbp]
\centering
\includegraphics[width=8cm]{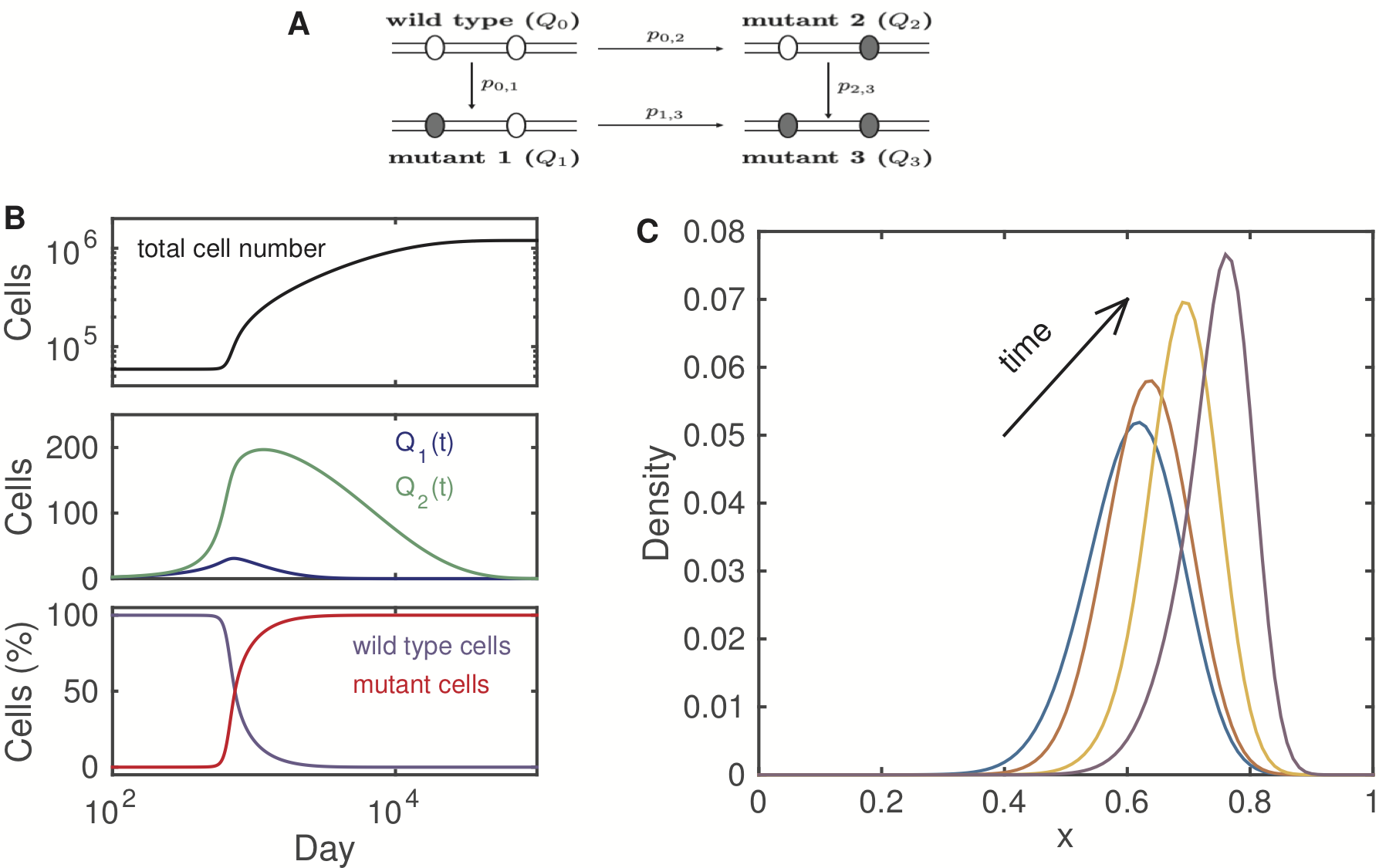}
\caption{\textbf{Tissue growth driven by cell heterogeneity and plasticity.} The blue line shows the simulated cell number, and the insets are scatter plots of the epigenetic states $(x_1, x_2)$ at different time points. For simulation details, refer to the Material and methods. The results were obtained using single-cell-based stochastic simulations.}
\label{fig:4}
\end{figure}

Figure \ref{fig:4} shows the stochastic simulation results from applying the proposed framework to a model of tissue growth driven by cell heterogeneity and plasticity. Here, there are two epigenetic states $\vec{x}=(x_1, x_2)$ that may affect the proliferation rate $\beta$ and the differentiation rate $\kappa$. We assumed that the inheritance of $x_1$ only depends on its own state at the mother cells, while the inheritance of $x_2$ depends on both $x_1$ and $x_2$ at the mother cell, then (where $\eta_i(\vec{y}) \equiv \eta$)
$$a_1(\vec{y}) = \eta \phi_1(y_1),\quad b_1(\vec{y}) = \eta (1 - \phi_1(y_1)),$$
$$a_2(\vec{y}) = \eta \phi_2(y_1, y_2),\quad b_2(\vec{y}) = \eta (1 - \phi_2(y_1, y_2)).$$
Here, we note that the relationships between epigenetic states may change following the developmental process or other conditions, and hence, the functions $\phi_i$ can depend on time $t$. In the simulations shown in Figure \ref{fig:4}, we took (refer to Material and methods for other functions and parameters)
$$\phi_1(y_1) = 0.08 + 1.0 \times \dfrac{(1.65 y_1)^{1.8}}{1 + (1.65 y_1)^{1.8}},$$
and
$$\phi_2(y_1, y_2) = 0.08 + \left(1 + f(t)\times\dfrac{0.4}{1 + (2.5 y_1)^6}\right) \times \dfrac{(1.56 y_2)^{1.8}}{1 + (1.56 y_2)^{1.8}},$$
where 
$$f(t) = \dfrac{1}{1 + e^{-(t-T_0)/1000}}$$
is a factor that specifies the time-dependent regulations. Figure \ref{fig:4} shows the dynamics of cell number and scatter plots of the epigenetic state of all cells at different time points. From the simulation, the increase in cell number from day 1500 to 2500 occurs together with the upregulation of the epigenetic state $x_2$ in the subpopulation of cells, which shows the cell heterogeneity and plasticity in tissue growth. 

\subsection{Modeling tumor development with cell-to-cell variance}
\label{sec:2.4}

As shown in Figure \ref{fig:2}F-G, to mimic the process of abnormal cell growth, we vary the differentiation rate and the inheritance probability. These variances to the model parameters can be a consequence of changes in the microenvironment that may affect all cells in the niche. Nevertheless, to model tumor development driven by gene mutations in individual cells, we need to modify the model equations to include changes in the genotypes.

\begin{figure}[htbp]
\centering
\includegraphics[width=10cm]{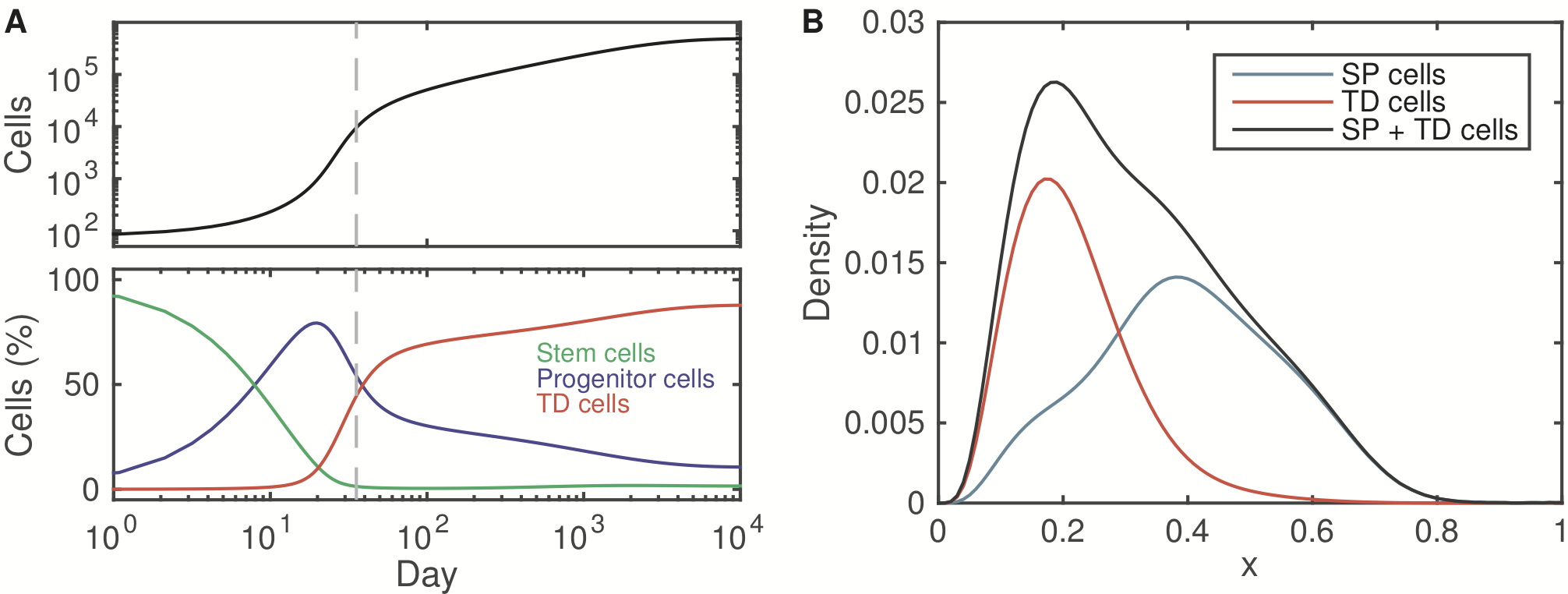}
\caption{\textbf{Simulated tumor development driven by mutations in proliferation and differentiation pathways.} \textbf{(A)}. Cell types and mutation probability $p_{i,j}$. Mutant 1 represents the cell type with an increased proliferation rate, mutant 2 represents the cell type with a decreased differentiation rate, and mutant 3 represents the cell type with a double mutation. \textbf{(B)}. Evolution dynamics of total cell counts (upper panel), mutant cells $Q_1(t)$ and $Q_2(t)$ (middle panel), and the fractions of wild-type and mutant cells (lower panel). \textbf{(C)}. The evolution of cell density during tumor development.}
\label{fig:5}
\end{figure}

To show the framework for modeling tumor development induced by driver gene mutations, we consider the process with two types of mutations, increasing the proliferation rate or decreasing the differentiation rate (Figure \ref{fig:5}A). Hence, let $Q_i(t,\vec{x})\ (i=0,1,2,3)$ represent the wild-type $(i=0)$ and the three mutant subpopulation $(i=1,2,3)$ cell counts, and $p_{i,j}(\vec{x})$ represents the mutation rates. Here, we note that the mutation rates may depend on the cell states. For simplicity, we assume that gene mutations occur only during cell division and that two daughter cells have the same mutation type. Therefore, Eq. \eqref{eq:2} can be extended as follows:
\begin{eqnarray*}
\dfrac{\partial Q_i(t,\vec{x})}{\partial t} &=&-Q_i(t,\vec{x}) (\beta_i(c,\vec{x}) + \kappa_i(\vec{x}))\\
&&{} + 2 \left(\int (1 - \sum_{j\not= i} p_{i,j}(\vec{y})) \beta_i(c_{\tau(\vec{y})},\vec{y})Q_i(t-\tau(\vec{y}),\vec{y})e^{-\mu(\vec{y})\tau(\vec{y})}p(\vec{x},\vec{y}) d \vec{y}\right)\nonumber\\
&&{}+ 2 \sum_{j\not=i} p_{j,i}(\vec{y})\int \beta_j(c_{\tau(\vec{y})},\vec{y})Q_j(t-\tau(\vec{y}),\vec{y})e^{-\mu(\vec{y})\tau(\vec{y})}p(\vec{x},\vec{y})d\vec{y},\nonumber \\
&&{} (1\leq i \leq 3)
\end{eqnarray*}
and
$$c(t) = \sum_{i=0}^3 \int Q_i(t,\vec{x}) \xi(\vec{x}) d \vec{x}.$$
Here, we consider only the driver mutation types, and at most one mutation occurs in each cell cycle, so that only the mutation rates $p_{0,1}, p_{0,2}$ and $p_{1,3}, p_{2,3}$ are nonzero values; otherwise, the mutation rates $p_{i,j}$ are zero (Figure \ref{fig:5}A). 

Figure \ref{fig:5}B-C shows the simulated dynamics. Single-mutation cells occurred prior to the obvious increase in the cell number, and the mutant cells eventually developed into double mutations and dominated the whole cell population (Figure \ref{fig:5}B). Moreover, our simulations suggest that stemness increases with evolutionary processes when we limited the mutations to proliferation and differentiation (Figure \ref{fig:5}C). Here, we considered only two types of mutations that often occur in the precancerous stage \cite{2015Natur.521...43D, Guo:2017ix}. To simulate a more complicated process of cancer development, we must extend the simulation to include more mutations, such as apoptosis, DNA damage repair, and immune response pathways.

\subsection{Modeling tissue growth with terminal differentiated cells}
In the abovementioned models, we considered only cells capable of self-renewal, \textit{e.g.}, stem cells and progenitor cells. Nevertheless, to model tissue growth, we must include terminal differentiated cells that lose the ability to progress through the cell cycle. Therefore, let $Q(t,\vec{x})$ represent cells with self-renewal ability as previously mentioned, and let $T(t,\vec{x})$ represent the number of terminally differentiated cells. The terminally differentiated cells are produced from the stem and progenitor cells with the rate $\kappa(\vec{x})$ and cleared with the rate $\nu(\vec{x})$. Hence, equation \eqref{eq:2} can be reformulated as follows:
\begin{equation}
\left\{
\begin{array}{rcl}
\dfrac{\partial Q(t,\vec{x})}{\partial t} &=& - Q(t,\vec{x}) (\beta(c,\vec{x}) + \kappa(\vec{x}))\\
&&{}\displaystyle + 2 \int \beta(c_{\tau(\vec{y})}, \vec{y}) Q(t - \tau(\vec{y}),\vec{y}) e^{-\mu(\vec{y})\tau(\vec{y})} p(\vec{x},\vec{y}) d \vec{y}\\
\dfrac{\partial T(t,\vec{x})}{\partial t} &=& \kappa(\vec{x}) Q(t,\vec{x}) - \nu(\vec{x}) T(t, \vec{x}).
\end{array}
\right.
\end{equation}
In the simulations shown in Figure \ref{fig:2}, when we considered the epigenetic state $0\leq x \leq 1$ as a stemness index and  distinguished the stem cells from progenitor cells with a boundary $x=x_0$, the numbers of stem cells, progenitor cells, and terminally differentiated cells can be given as follows:
$$Q_{\mathrm{stem}}(t) = \int_{x_0}^1 Q(t,x) d x,\ Q_{\mathrm{progenitor}}(t)=\int_0^{x_0}Q(t,x) d x,\ T(t)=\int_0^1T(t,x)dx.$$

\begin{figure}[htbp]
\centering
\includegraphics[width=10cm]{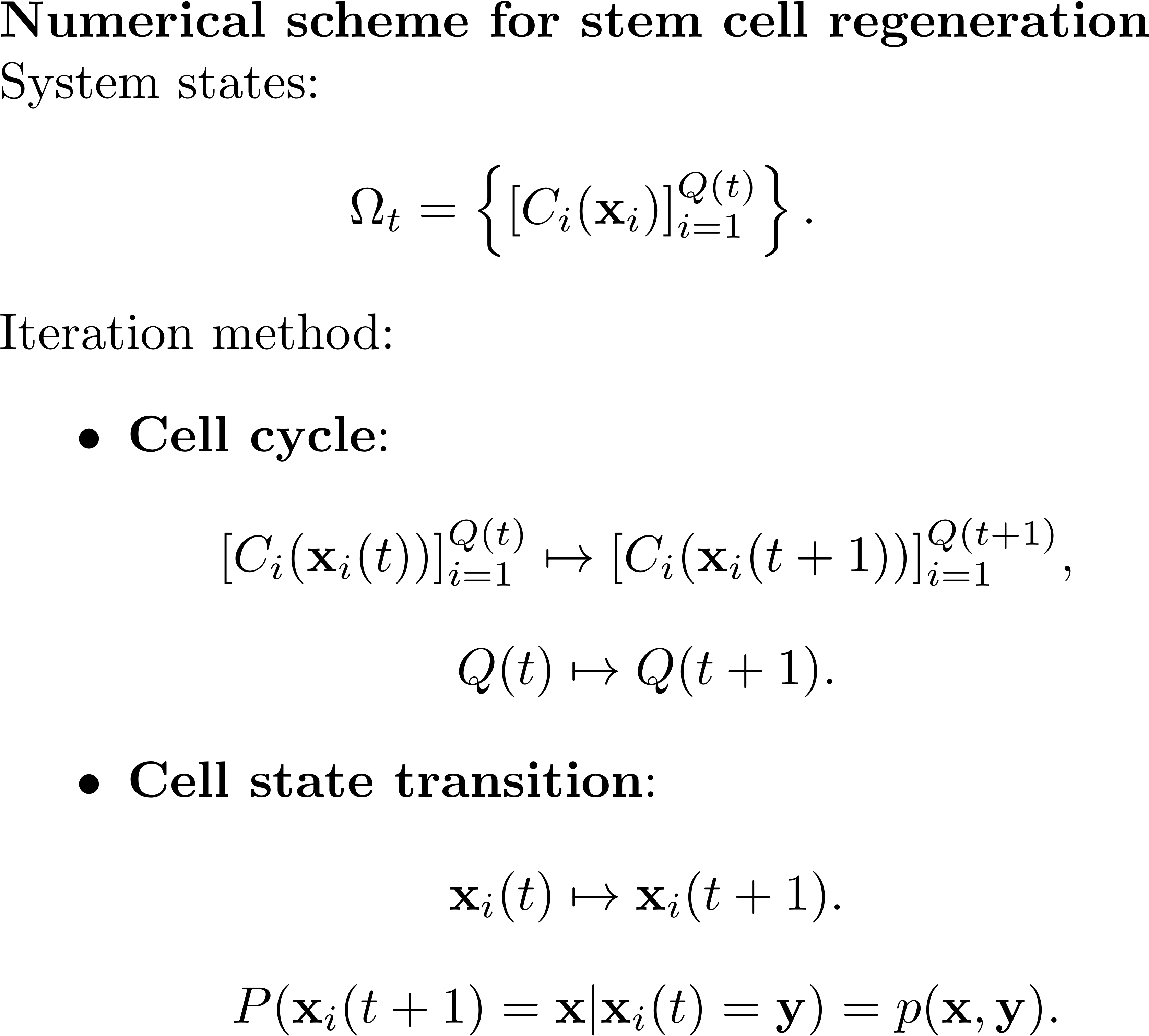}
\caption{\textbf{The dynamics of tissue growth.} \textbf{(A).} Simulated dynamics of tissue growth beginning with $100$ stem cells at day $0$. Lower panel: fractions of stem cells, progenitor cells, and terminally differentiated cells. \textbf{(B).} Density of the epigenetic state of the three cell populations at day $35$ (gray dashed line shown in (A)). Here, SP cells indicate stem and progenitor cells, and TD cells indicate terminally differentiated cells. Stem cells were defined as $0.7 \leq x \leq 1$, and progenitor cells were defined as $0<x<0.7$. See the Material and methods for the simulation details.}
\label{fig:6}
\end{figure}

This equation provides a model of multistage cell lineages shown in previous studies \cite{Adimy:2014gs, Gaspari:2018kl, Lander:2009fr}. The tissue size is given by the total cell number as follows:
$$S(t)=Q_{\mathrm{stem}}(t)+Q_{\mathrm{progenitor}}(t)+T(t),$$
and the distribution of stemness among all tissue cells is given by 
$$f(t,x)=\dfrac{Q(t,x)+T(t,x)}{S(t)}.$$
Figure \ref{fig:6} shows the simulated dynamics beginning with $100$ stem cells, which reveal the transition between stem and progenitor cells and the differentiation to terminally differentiated cells. Figure \ref{fig:6}B shows the density of phenotypically distinct cell populations among the stem cells, progenitor cells and the terminally differentiated cells. 

\section{Discussion}
   
Regeneration of stem and progenitor cells is a basic cellular behavior associated with development, aging, and many complex diseases in multicellular organisms. In this study, to overlook the genetic details, we established a general mathematical framework to describe the process of stem and progenitor cell regeneration. This framework highlights cell heterogeneity and connects heterogeneity with cellular behaviors, \textit{e.g.}, proliferation, apoptosis, and differentiation/senescence. Cell heterogeneity is often associated with epigenomic markers that are subject to stochastic inheritance during cell division and is described through an inheritance probability function. Hence, the framework provides a multiscale model that incorporates microscopic epigenetic states and gene expressions with macroscopic tissue growth through mesoscopic cell behaviors. We believe that this formula is helpful in answering the Weinberg question \cite{2007Natur.449..978W}. Despite the generality of this formula, different assumptions regarding the kinetic rate functions and the inheritance probability can be applied to describe various biological processes related to stem cell regeneration (Figure \ref{fig:2}).

In the proposed framework, all stem and progenitor cells are described with a single compartment model, and different phenotypic cells are not distinguished explicitly. This approach differs from the differentiation tree models that are widely used to describe the maintenance of hierarchically organized tissues. Recently, an increasing number of experiments have shown a continuous spectrum of cell differentiation, which challenges the distinction between stem and progenitor cell populations \cite{2018Natur.553..418L, Nestorowa:2016iq, Velten:2017jj}. Stochastic state transitions between different phenotypic cells lead to a dynamic equilibrium among a population of self-renewing cells \cite{Gupta:2010fk, 2016PNAS..113.2672L, Santos:2018kj}. Our model suggests that discrimination between cell types may not be necessary to describe tissue homeostasis. Different subtypes of cells can be characterized by their kinetic rates of proliferation, apoptosis, differentiation, senescence, \textit{etc}. For convenience, these dynamic features can be referred to as the \textit{kinotype}, which is analogous to the genotype, epigenotype, or phenotype of a cell. The kinotype of cells is often associated with the activities of specific genes enriched in the related pathways. If the relationships between kinetic rates and the expression of these genes are known, the proposed framework can be extended to include the roles of specific genes. Therefore, in the future, we can apply the framework to develop  predictable models to investigate how variations in specific genes alter the long-term dynamics of tissue growth.

Although the probabilistic epigenetic inheritance is considered, Eq. \eqref{eq:2} is a deterministic equation that describes the dynamics of cell densities with different epigenetic states. This model often provides information regarding the average of multiple cells. To model a single-cell stochastic behaviors, we need to perform stochastic simulations that explicitly account for random events. Eq. \eqref{eq:2} suggests a numerical scheme of multiscale modeling for tissue growth where a multiple cell system is represented by a collection of epigenetic states in each cell. In each time interval, each cell undergoes proliferation, apoptosis, or terminal differentiation with a probability given by the corresponding kinetic rates, so that the cell number changes accordingly. Moreover, for each cell undergoing cell division, the mother cell is replaced by two daughter cells, and the epigenetic state of each daughter cell is determined according to the predefined inheritance probability. An example of a single-cell-based stochastic simulation is shown in Figure \ref{fig:4}. In our previous study, this computational scheme was applied to model the process of inflammation-induced tumorigenesis. Simulation results reproduced the two-stage tumorigenesis dynamics and revealed the competing oncogenic and onco-protective roles of inflammation. Based on the simulated evolution of single-cell states, we were able to uncover the detailed process of cancer development. We expect that further development of the model framework is capable of modeling the evolutionary dynamics of cancer.

\section*{Material and methods}
\subsection*{\textbf{Resource}}

Source MATLAB and \texttt{C++} codes for the study are available from\\ \texttt{https://github.com/jzlei/StemCell}.

\subsection*{\textbf{Age-structured model and delay differential equation models}}

In the G0 cell cycle model, $Q(t)$ is the number of resting-phase stem cells, $s(t,a)$ is an age-structured quantify that represents the population of proliferating stem cells, and the age $a=0$ is their time of entry into the proliferative state. The resting-phase cells can either reenter the proliferative phase at a rate $\beta(Q)$ or differentiate into downstream cell lines at a rate $\kappa$. The proliferating stem cells are assumed to undergo mitosis at a fixed time $\tau$ after entry into the proliferating compartment and to be lost randomly at a rate $\mu$ during the proliferating phase. Each normal cell generates two resting-phase cells at the end of mitosis. Here, the units of the cell population are often measured by the number of cells per unit body weight, \textit{e.g.}, $\mathrm{cells/kg}$, and the units for the rates of proliferation, differentiation, and apoptosis are $\mathrm{day}^{-1}$. 

The above assumptions yield the following partial differential equations \cite{Jinzhi:2011bl}:
\begin{equation}
\label{eq:8}
\begin{array}{lcl}
\nabla s(t,a) = -\mu s(t,a),\quad & (t>0, 0<a<\tau)\\
\dfrac{d Q}{d t} = 2 s(t,\tau)  - (\beta(Q) + \kappa) Q, \quad & (t>0)
\end{array}
\end{equation}
Here, $\nabla = \partial/\partial t + \partial/\partial a$. The boundary condition at $a=0$ is as follows:
\begin{equation}
\label{eq:9}
s(t,0) = \beta(Q(t)) Q(t),
\end{equation}
and the initial conditions are
\begin{equation}
\label{eq:10}
s(0,a) = g(a),\quad (0\leq a \leq \tau);\quad Q(0) = Q_0.
\end{equation}
Eqs. \eqref{eq:8}-\eqref{eq:10} provide a general age-structured model of homogeneous stem cell regeneration.

Integrating Eqs. \eqref{eq:8}-\eqref{eq:9} with the characteristic line method, we obtain the following close-form differential equation \cite{Jinzhi:2011bl}:
\begin{equation}
\label{eq:11}
\dfrac{d Q}{d t} = - (\beta(Q) + \kappa) Q + \left\{
\begin{array}{lcl}
2 g(\tau - t) e^{-\mu t}, \quad & 0 < t \leq \tau\\
2 e^{-\mu \tau} \beta(Q_\tau)Q_\tau, & t > \tau
\end{array}
\right.
\end{equation}
where $Q_\tau = Q(t-\tau)$. When we only consider the long-term behavior $(t > \tau)$ and shifted the original time point to $\tau$, we obtained the delay differential equation model
\begin{equation}
\label{eq:12}
\dfrac{d Q}{d t} = - (\beta(Q) + \kappa) Q + 2 e^{-\mu \tau} \beta(Q_\tau) Q_{\tau}.
\end{equation}
This equation describes the general population dynamics of homogeneous stem cell regeneration.

\subsection*{\textbf{Formulation of the proliferation rate}}

The effect of feedback regulation from the cell population to the proliferation rate is given by the function $\beta(Q)$. Biologically, the self-renewal ability of a cell is determined by both microenvironmental conditions, \textit{e.g.}, growth factors and various types of cytokines, and intracellular signaling pathways, \textit{e.g.}, growth factor receptors and cell cycle checkpoints, such as fibroblast growth factors (FGFs) and the transforming growth factor beta (TGF-$\beta$) family \cite{Massague:2012bd,Nakao:1997dm,Ornitz:2001vh}. The exact activation pathways that regulate the self-renewal of stem cells are poorly understood. Here, we derive a phenomenological formulation based on simple but general assumptions.

There are positive and negative signals for stem cell proliferation. We assumed that positive growth factors are secreted by the niche and that growth factor inhibitors are released by the cells. Different types of cytokines bind to cell surface receptors to regulate the cell behavior. Let $[\mathrm{L}]$ denote the concentration of ligands for growth factor inhibitor;  $[\mathrm{R}]$ denote the density of free receptors; $[\mathrm{R}^*]$ denote the density of activated receptors; and $Q$ denote the stem cell number. The total number of receptors is
\begin{equation}
\label{eq:s3}
[\mathrm{R}]+[\mathrm{R}^* ] = m Q,
\end{equation}
where $m$ is the average number of receptors per cell. If $n$ ligands are required to activate one receptor, we assumed that the ligands bind to the receptor following the law of mass action as follows:
$$[\mathrm{R}] + n [\mathrm{L}] \leftrightarrows [\mathrm{R}^* ].$$
At equilibrium, we have the following equation:
\begin{equation}
\label{eq:13}
[\mathrm{R}] [\mathrm{L}]^n = K [\mathrm{R}^*]
\end{equation}
where $K$ is the equilibrium constant. We further assume that the activated receptors inhibit cell proliferation so that the proliferation rate $\beta$ is proportional to the fraction of free receptors on a cell,
\begin{equation}
\label{eq:14}
\beta =\beta_0\dfrac{[\mathrm{R}]}{mQ}.
\end{equation}
From \eqref{eq:s3}-\eqref{eq:13}, we obtain
\begin{equation}
\label{eq:s14}
\dfrac{[\mathrm{R}]}{m Q} = \dfrac{K}{K + [\mathrm{L}]^n}.
\end{equation}

When the ligands are secreted from stem cells and are cleared at a constant rate, the ligand concentration is proportional to the cell number, which gives $[\mathrm{L}]=\sigma Q$. Thus, Eqs. \eqref{eq:14} and \eqref{eq:s14} yield the final form of the proliferation rate
\begin{equation}
\label{eq:15}
\beta(Q)=\beta_0  \dfrac{\theta^n}{\theta^n+Q^n},
\end{equation}
where $\theta = \sqrt[n]{K}/\sigma$ is the $50\%$ effective coefficient (EC50).

From Eq. \eqref{eq:15}, the proliferation rate, which is important for tissue homeostasis, approaches $0$ due to the antiproliferation signals when the cell number $Q$ is sufficiently large. However, the capability of self-sufficiency to growth signals and insensitivity to antigrowth signals are the two characteristics of cancer that enable malignant tumor cells to escape antigrowth signals \cite{Hanahan:2000hx}. Hence, to model tumor development, in addition to possible changes in parameters (increasing $\theta$ or $\beta_0$), the proliferation rate can also be modified to include a nonzero constant $\beta_1$ for self-sustained growth signals as follows:
\begin{equation}
\label{eq:16}
\beta(Q)=\beta_0  \dfrac{\theta^n}{\theta^n+Q^n} + \beta_1.
\end{equation}
For a more realistic model, cell growth can be restricted by physical interactions, and hence $\beta_1$ may decrease to zero when the cell number $Q$ becomes large enough.

\subsection*{\textbf{Steady state of the G0 cell cycle model and oncogenic signaling pathways}} 

From Eq. \eqref{eq:12}, the steady state $Q(t)\equiv Q^*$ is given by the equation
$$
-(\beta(Q^* )+\kappa) Q^*+2e^{-\mu \tau} \beta(Q^*) Q^*=0,
$$
which yields either $Q^*=0$, or
\begin{equation}
\label{eq:17}
\beta(Q^* )=\dfrac{\kappa}{2e^{-\mu \tau}-1}.
\end{equation}
When $\beta(Q)$ is given by Eq. \eqref{eq:16}, Eq. \eqref{eq:12} has a unique positive steady state if and only if 
$$\beta_0 > \dfrac{\kappa}{2e^{-\mu \tau}-1} - \beta_1 > 0.$$ 
In particular, when
\begin{equation}
\label{eq:18}
\beta_1\geq \dfrac{\kappa}{2e^{-\mu \tau}-1},
\end{equation}
Eq. \eqref{eq:12} has only a zero steady state, and the zero solution is unstable. Hence, all positive solutions approach infinity, which corresponds to uncontrolled growth. Therefore, inequality \eqref{eq:18} summarizes a general condition for uncontrolled growth, \textit{i.e.}, malignant tumors. Biologically, Eq. \eqref{eq:18} is satisfied if there is self-sufficiency in growth signals and/or insensitivity to antigrowth signals (increasing $\beta_1$), evasion of apoptosis (decreasing $\mu$), and dysregulation in the differentiation and/or senescence pathways (decreasing $\kappa$), which are well-known hallmarks of cancer \cite{Hanahan:2000hx}.

\subsection*{\textbf{Age-structured model of heterogeneous stem cell regeneration}}

When stem cell heterogeneity is considered and we assume that the apoptosis rates of cells during cell division are dependent on the epigenetic states of cells before entering the proliferating phase, the age-structured model (Eq. \eqref{eq:8}) becomes
\begin{eqnarray*}
\nabla s(t,a, \vec{x}) &=&  -\mu(\vec{x}) s(t,a,\vec{x}),\quad (t>0, 0<a < \tau(\vec{x}))\\
\dfrac{\partial Q(t,\vec{x})}{\partial t} &=& 2 \int s(t, \tau(\vec{y}), \vec{y}) p(\vec{x},\vec{y}) d \vec{y} - (\beta(c(t), \vec{x}) + \kappa(\vec{x})) Q(t,\vec{x}),\quad (t>0)
\end{eqnarray*}
and
$$s(t,0,\vec{x})=\beta(c(t), \vec{x})Q(t, \vec{x}).$$
Here, $\vec{x}$ represents the epigenetic states of the cells. In the first equations, when we consider the epigenetic state $\vec{x}$ as the parameters, the characteristic line method remains valid, which gives the following equation (here we only show the result for long-term behavior):
$$s(t,\tau(\vec{x}),\vec{x})=\beta(c(t-\tau(\vec{x}),\vec{x}),\vec{x})Q(t-\tau(\vec{x}),\vec{x}) e^{-\mu(\vec{x})\tau(\vec{x})}.$$
Substituting $s(t,\tau(\vec{x}),\vec{x})$ into the second equation, we obtain 
\begin{eqnarray}
\label{eq:19}
\dfrac{\partial Q(t,\vec{x})}{\partial t}  &=& -Q(t,\vec{x})(\beta(c,\vec{x})+\kappa(\vec{x}))\\
&&{}+2\int \beta(c_{\tau(\vec{y})},\vec{y})Q(t-\tau(\vec{y}),\vec{y}) e^{-\mu(\vec{y})\tau(\vec{y})}) p(\vec{x},\vec{y})d\vec{y}, \nonumber
\end{eqnarray}
which gives Eq. \eqref{eq:2} for heterogeneous stem cell regeneration.

Let
$$
Q(t) = \int Q(t,\vec{x}) d \vec{x},
$$
which is the total cell number, and integrate Eq. \eqref{eq:19} with $\vec{x}$; we then obtain the following equation (here we note $p(\vec{x},\vec{y})d\vec{x}=1$)
$$
\dfrac{dQ}{dt}=-\int Q(t,\vec{x}) (\beta(c,\vec{x}) + \kappa(\vec{x})) d \vec{x} + 2 \int \beta(c_{\tau(\vec{x})},\vec{x}) Q(t-\tau(\vec{x}), \vec{x}) e^{-\mu(\vec{x}) \tau(\vec{x})} d \vec{x}.
$$
Define
$$f(t, \vec{x})= \dfrac{Q(t, \vec{x})}{Q(t)} $$
as the density of cells with a given epigenetic state $\vec{x}$, then
\begin{eqnarray*}
\dfrac{\partial f(t,\vec{x})}{\partial t} &=&\dfrac{1}{Q(t)} \dfrac{\partial Q(t,\vec{x})}{\partial t} - \dfrac{Q(t,\vec{x})}{Q(t)^2} \dfrac{d Q(t)}{d t}\\
&=& - f(t,\vec{x}) (\beta(c, \vec{x}) + \kappa(\vec{x}))\\
&&{} + \dfrac{2}{Q(t)}\int \beta(c_{\tau(\vec{y})},\vec{y})Q(t-\tau(\vec{y}),\vec{y})e^{-\mu(\vec{y})\tau(\vec{y})}p(\vec{x},\vec{y})d\vec{y}\\
&&{}-\dfrac{f(t,\vec{x})}{Q(t)}\Big(-\int Q(t,\vec{x}) (\beta(c,\vec{x}) + \kappa(\vec{x}))d\vec{x}\\
&&{}\qquad\qquad + 2 \int \beta(c_{\tau(\vec{x})},\vec{x}) Q(t-\tau(\vec{x}),\vec{x})e^{-\mu(\vec{x})\tau(\vec{x})}d\vec{x}\Big)\\
&=& - f(t,\vec{x}) \int f(t,\vec{y}) (\beta(c,\vec{x})  + \kappa(\vec{x})) d \vec{y}\\
&&{} + f(t,\vec{x}) \int f(t,\vec{y}) (\beta(c,\vec{y}) + \kappa(\vec{y}))d\vec{y}\\
&&{} + \dfrac{2}{Q(t)}\int \beta(c_{\tau(\vec{y})},\vec{y})Q(t-\tau(\vec{y}),\vec{y})e^{-\mu(\vec{y})\tau(\vec{y})}p(\vec{x},\vec{y})d\vec{y}\\
&&{}-\dfrac{2}{Q(t)}\int \beta(c_{\tau(\vec{y})},\vec{y})Q(t-\tau(\vec{y}),\vec{y})e^{-\mu(\vec{y})\tau(\vec{y})}f(t,\vec{x})d\vec{y}\\
&=&-f(t,\vec{x})\int f(t,\vec{y}) \left((\beta(c,\vec{x}) + \kappa(\vec{x})) - (\beta(c,\vec{y}) + \kappa(\vec{y}))\right)d\vec{y}\\
&&{}+\dfrac{2}{Q(t)} \int \beta(c_{\tau(\vec{y})},\vec{y})Q(t-\tau(\vec{y}),\vec{y})e^{-\mu(\vec{y})\tau(\vec{y})}(p(\vec{x},\vec{y}) - f(t,\vec{x}))d \vec{y}.
\end{eqnarray*}
Hence, we have the equation
\begin{eqnarray}
\label{eq:20}
\dfrac{\partial f(t,\vec{x})}{\partial t} &=& -f(t,\vec{x})\int f(t,\vec{y}) \left((\beta(c,\vec{x}) + \kappa(\vec{x})) - (\beta(c,\vec{y}) + \kappa(\vec{y}))\right)d\vec{y}\\
&&{}+\dfrac{2}{Q(t)} \int \beta(c_{\tau(\vec{y})},\vec{y})Q(t-\tau(\vec{y}),\vec{y})e^{-\mu(\vec{y})\tau(\vec{y})}(p(\vec{x},\vec{y}) - f(t,\vec{x}))d \vec{y}.\nonumber 
\end{eqnarray}

\subsection*{\textbf{The inheritance probability $p(\vec{x},\vec{y})$}}

In Eq. \eqref{eq:2}, the inheritance probability $p(\vec{x},\vec{y})$ is essential in describing the heterogeneity of cells. The function $p(\vec{x},\vec{y})$ is associated with the process of cell division during which the epigenetic code and molecules are distributed to daughter cells through complex regulation mechanisms that are not well understood, as well as stochastic gene expression. Hence, the general mathematical formula of the function $p(\vec{x},\vec{y})$ remains unknown. Here, we propose an attempt to define the function based on the random inheritance of histone modifications.

In eukaryotic cells, most DNA sequences are enclosed in nucleosomes in which DNA sequences wrap around a histone octamer that is composed of one $\mbox{(H3-H4)}_2$ tetramer capped by two H2A-H2B dimers. These histones can undergo diverse posttranslational covalent modifications that lead to either active or repressive gene expression activity \cite{Bintu:2016dv,Kouzarides:2007js,Lachner:2003gt}. The patterns of histone modification dynamically change over time and hence define a dynamic histone code for the transcription activity. The dynamics of histone modifications consist of complex processes, including nucleosome assembly, writing and erasing of modification markers, and random inheritance during DNA replication \cite{Probst:2009iq,SerraCardona:2018bz}. Detailed computational models for the process of histone modification and random inheritance over the cell cycle remain a challenging issue in computational biology. When we consider the main process of writing and erasing the modification markers that are modulated by the related enzymes, the kinetics of histone modification can be modeled through stochastic simulations \cite{Huang:2017jr,Ku:2013kx}. 

In a proposed dynamic model of histone modification \cite{Huang:2017jr,Ku:2013kx}, bivalent modifications of histone H3, trimethylation of H3 lysine 4 (H3K4me3) and the trimethylation of H3 lysine 27 (H3K27me3) are considered. Each H3 histone can be in one of the following states: unmodified (U), modified by the activating marker H3K4me3 (A), or modified by the repressing marker H3K27me3 (R). Each nucleosome can be in one of six physical nucleosome states, UU, AU, UR, AA, AR, or RR.  The nucleosome states dynamically change according to methylation/demethylation, which are regulated by the corresponding enzymes. During DNA replication, parental histones and newly synthesized histones are randomly distributed on daughter strands. To avoid the dilution of histone markers, maintenance modifications in the new histones can be achieved by using a neighboring histone as a template \cite{Probst:2009iq}. Hence, writing enzyme activities are dependent on the states of neighboring nucleosomes. Thus, changes in the nucleosome state over the cell cycle due to the random distribution of histone markers during DNA replication and kinetic methylation/demethylation can be tracked with stochastic simulations \cite{Huang:2017jr}.

Based on the abovementioned model simulation, we are able to study how the nucleosome states of daughter cells depend on their mother cells. For example, considering a DNA segment with $N$ nucleosomes, we counted the number ($N_A$) of nucleosomes with active markers (either AA or AU) at each cycle. The simulation results suggested that considering the state of the mother cell, the active nucleosome number is a random number with a binomial distribution with the parameter (success probability $p$) dependent on the state of the mother cell as follows \cite{Huang:2017jr}:
$$
P(N_{A,k+1} = n | N_{A,k} = m) = C_N^n p^n (1-p)^{N-n}, \quad \mathrm{where}\ p = p(m).
$$
Considering the nucleosome state through the fraction ($f_A=N_A/N$) of active nucleosome numbers, the binomial distribution can be extended to a continuous probability distribution defined on the interval $[0,1]$, which is given by the beta distribution:
$$P(f_{A,k+1} = x | f_{A,k} = y) = \dfrac{x^{a(y)-1} (1-x)^{b(y)-1}}{B(a(y),b(y))}.$$
Hence, for the specific situation of the random inheritance of histone modifications during the cell cycle, we can use the beta distribution probability as the inheritance probability function $p(x,y)$. Here, we extend the above formulation to general cases and propose Eq. \eqref{eq:5-0} as the inheritance functions.  

\subsection*{\textbf{Beta distribution}}

The beta distribution is one of the few common distributions defined on a finite interval and is parameterized by two positive shape parameters that appear as positive exponents of the random variable. As the shape parameters vary, the beta distribution can take different shapes, including strictly decreasing ($a\leq 1, b> 1$), strictly increasing ($a>1, b\leq 1$), U-shaped ($a<1, b<1$), or unimodal ($a>1, b>1$). The probability density function (PDF) of the beta distribution is given by a power function of the variable $x$ and its reflection $(1-x)$, 
$$
f(x;a,b)=\dfrac{x^{a-1} (1-x)^{b-1}}{B(a,b)}, \quad B(a,b)=\dfrac{\Gamma(a)\Gamma(b)}{\Gamma(a+b)},
$$
where $\Gamma(z)$ is the gamma function.

For a random variable $X$ beta-distributed with parameters $a$ and $b$, which is denoted by $X\sim \mathrm{beta}(a,b)$, the mean and variance are, respectively,
$$\mathrm{E}[X]=\dfrac{a}{a+b},\quad \mathrm{Var}[X]=\dfrac{ab}{(a+b)^2 (a+b+1)}.$$
Then, it is easy to obtain
$$\mathrm{Var}[X]=\dfrac{1}{1+a+b}\mathrm{E}[X](1-\mathrm{E}[X]).$$
Hence, if we assume
$$\mathrm{E}[X]=\phi,\quad \mathrm{Var}[X]=\dfrac{1}{1+\eta}\phi (1-\phi),$$
then
$$\phi = \dfrac{a}{\eta},\ \eta=a+b,$$
which gives
$$a=\eta \phi,\ b = \eta (1-\phi).$$
This gives Eq. \eqref{eq:5} that determines the shape parameters from the functions $\phi_i(\vec{y})$ and $\eta_i(\vec{y})$.

\subsection*{\textbf{Simulations for stem cell regeneration}}

Here, we present a simple example to show the numerical scheme for simulating stem cell regeneration based on the proposed model equations. 

We consider a situation with one epigenetic state $x$ ($0\leq x \leq1$), the stemness index, that affects only cell proliferation and differentiation so that only the rates $\beta$ and $\kappa$ are dependent on the epigenetic state $x$. Therefore, we have the following model equation:
$$
\dfrac{\partial Q(t,x)}{\partial t} = - Q(t,x) (\beta(\textcolor{red}{c}(t), x) + \kappa(x)) + 2 e^{-\mu \tau} \int_0^1 \beta(c(t-\tau),y) Q(t-\tau,y) p(x,y) d y.
$$
Here, $\xi(x)=1$ so that
$$c(t) = \int_0^1 Q(t,x) d x.$$

To specify the rate functions $\beta$ and $\kappa$, we assume that the state $x$ affects the proliferation and differentiation rates in a manner similar to the stemness so that a large value of $x$ indicates stem cells with a low proliferation rate, an intermediate value of $x$ indicates progenitor cells with a high proliferation rate, and the terminated differentiation rate $\kappa$ is a decreasing function of $x$ and approaches zero when $x$ is large. These functions are mathematically expressed as
$$
\beta(c,x) = \beta_0(x)\dfrac{\theta}{\theta+c},\ \beta_0(x) = \bar{\beta}\times \dfrac{a_1 x + (a_2 x)^6}{1 + (a_3 x)^6},$$
and
$$\kappa(x) = \kappa_0 \times \dfrac{1}{1 + (b_1 x)^6}.$$

The inheritance probability $p(x,y)$ is defined from the beta distribution density function through predefined functions $\phi(y)$ and $\eta(y)$ as follows:
$$
p(x,y) = \dfrac{x^{a(y)-1} (1-x)^{b(y)-1}}{B(a(y), b(y))},\ B(a,b) = \dfrac{\Gamma(a) \Gamma(b)}{\Gamma(a+b)},
$$
and
$$a(y) = \eta(y) \phi(y),\ b(y) = \eta(y) (1-\phi(y)).$$

\begin{figure}[htbp]
\centering
\includegraphics[width=10cm]{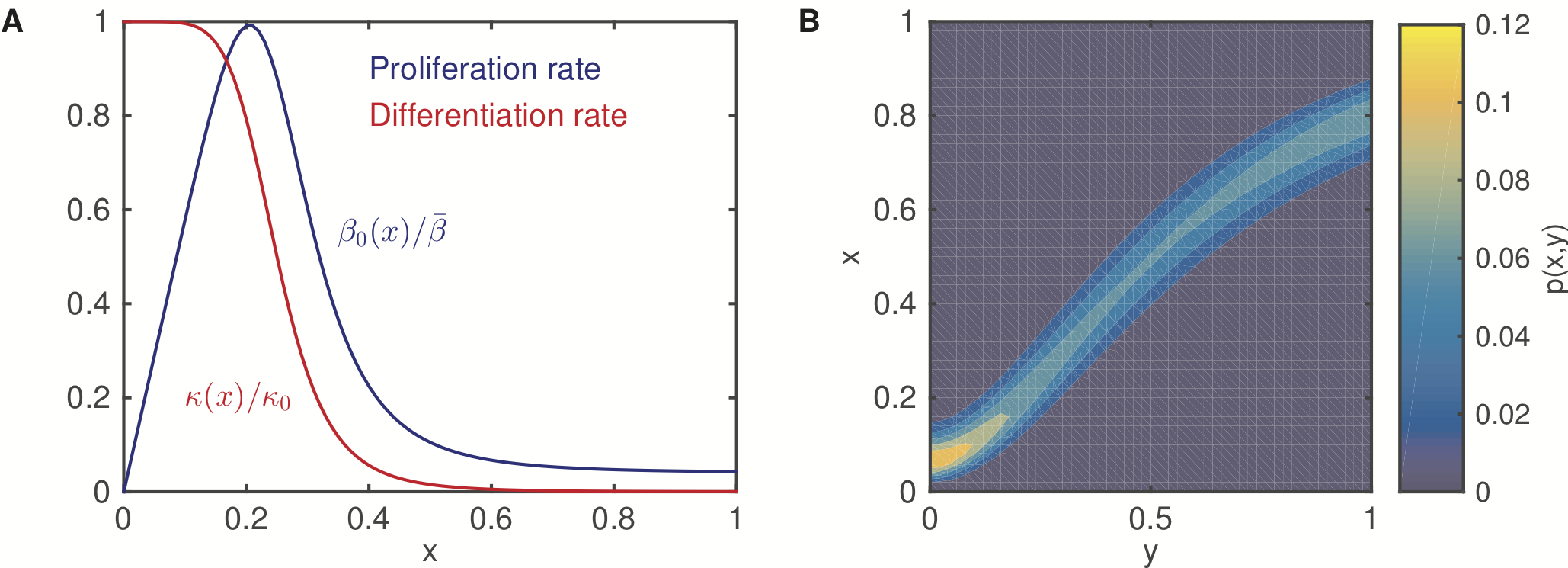}
\caption{\textbf{Example of the rate functions.} (A). The proliferation rate $\beta_0(x)$ and the differentiation rate $\kappa(x)$. (B). The inheritance probability $p(x,y)$. Here, the parameters are $a_1=5.8, a_2=2.2, a_3=3.75, b_1=4.0$, and $\phi(x)=0.08+1.0\times\frac{(1.65 x)^{1.8}}{1 + (1.65 x)^{1.8}}, \eta(x)=60$.}
\label{fig:7}
\end{figure}

Figure \ref{fig:7} shows the functions $\beta_0(x)$, $\kappa(x)$, and $p(x,y)$.

In the simulations shown in Figure \ref{fig:2}, the parameters are set as follows: $\mu = 2.0\times 10^{-4}\, \mathrm{h}^{-1}, \tau = 20\, \mathrm{h}, \theta = 10^3\mathrm{cells}, a_1 = 5.8, a_2=2.2, a_3=3.75, b_1=4.0, \eta(x)=60$, and
\begin{itemize}
\item in (B)-(C): $\bar{\beta} = 0.12\, \mathrm{h}^{-1}, \kappa_0 = 0.02\,  \mathrm{h} ^{-1}, \phi(x) = 0.08 +1.0\times\frac{(1.65 x)^{1.8}}{1 + (1.65 x)^{1.8}}$;
\item in (D)-(E): $\bar{\beta} = 0.12\, \mathrm{h}^{-1}, \kappa_0 = 0.02\, \mathrm{h} ^{-1}, \phi(x) = 0.04 +0.9\times\frac{(1.65 x)^{1.8}}{1 + (1.65 x)^{1.8}}$;
\item in (F)-(G): $\bar{\beta} = 0.12\, \mathrm{h}^{-1}, \kappa_0 = 0.0002\,  \mathrm{h}^{-1}, \phi(x) = 0.08 +1.2\times\frac{(1.65 x)^{1.8}}{1 + (1.65 x)^{1.8}}$.
\end{itemize}

In the simulation shown in Figure \ref{fig:3}, $\bar{\beta}=4.8 \mathrm{h}^{-1}, \kappa_0=0.4 \mathrm{h}^{-1}$, and the other parameters are the same as those shown in Figure \ref{fig:2}B.

In the simulation shown in Figure \ref{fig:5}, we set the wild-type cell parameters according to those shown in Figure \ref{fig:2}B, and $p_{0,1}=p_{0,2}=0.5\times 10^{-4}, p_{1,3} =p_{2,3} =1\times 10^{-4}$. For mutation 1, the proliferation rate is twice that of the wild-type cells, and for mutation 2, the differentiation rate is $1/10$ that of the wild-type cells. 

In the simulation shown in Figure \ref{fig:6}, we include the terminally differentiated cells into the model, and solve the equation
$$
\left\{
\begin{array}{rcl}
\displaystyle\dfrac{\partial Q(t,x)}{\partial t} &=& \displaystyle - Q(t,x) (\beta(c(t), x) + \kappa(t,x)) \\
&&{} \displaystyle + 2 e^{-\mu \tau} \int_0^1 \beta(c(t-\tau),y) Q(t-\tau,y) p(x,y) d y\\
\displaystyle \dfrac{\partial T(t,x)}{\partial t} &=& \displaystyle \kappa(t,x) Q(t,x) - \nu(x) T(t,x).
\end{array}
\right.
$$
Here, we set the death rate of terminally differentiated cells $\nu(x) = 2\times 10^{-5}\, \mathrm{h}^{-1}$, the differentiation rate increases from $0$ to a normal level following 
$$\kappa(t,x)=\kappa_0 \times \dfrac{1}{1+e^{-(t-T_0)/1000}}\times \dfrac{1}{1 + (b_1 x)^6},$$ 
with $\kappa_0 = 0.02\, \mathrm{h}^{-1}, T_0 = 1680\, \mathrm{h} (=70\,\mathrm{day})$, and the other parameters are the same as those shown in Figure \ref{fig:2}B.

To determine the model parameters, the kinetic rates of cell proliferation, apoptosis, and terminal differentiation can be obtained from experiments of cell growth and differentiation. Parameters in the inheritance functions can be determined by fitting the simulation results with single-cell sequencing data, such as the distributions of gene expressions of each gene. Further discussions of parameter determination are beyond the scope of this paper and depend on the specific applications for which they are needed. 

\subsection*{\textbf{Modeling tissue growth driven by cell heterogeneity and plasticity}}
Here, we show an application of the model framework to a more complex process with tissue growth driven by cell heterogeneity and plasticity. The simulation results are shown in Figure \ref{fig:4}.

We assume that there are two epigenetic states $\vec{x} = (x_1, x_2)$, where $x_1$, similar to the above stemness index, mainly affects the proliferation rate $\beta$ and the differentiation rate $\kappa$, and $x_2$ affects the efficiency of the growth factor inhibitors so that the parameter $\theta$ is a function of $x_2$. Thus, we have the equation
\begin{eqnarray}
\label{eq:s2}
\dfrac{\partial Q(t, \vec{x})}{\partial t} &=& - Q(t, \vec{x}) \left(\beta(c(t), \vec{x}) + \kappa(\vec{x})\right)\\
&&{} + 2 e^{-\mu \tau} \int_\Omega \beta(c(t-\tau), \vec{y}) Q(t-\tau, \vec{y}) p(\vec{x}, \vec{y}) d \vec{y}, \nonumber
\end{eqnarray}
where $\Omega = [0,1]\times [0,1]$, and
$$c(t) = \int_\Omega Q(t, \vec{x}) d \vec{x}.$$
Similar to the above model, we write
$$
\beta(c, \vec{x}) = \beta_0(x_1) \dfrac{\theta(x_2)}{\theta(x_2) + c},\quad \kappa(\vec{x}) = \kappa_0\times \dfrac{1}{1 + (b_1 x_1)^6},$$
where
$$\beta_0(x_1) = \bar{\beta} \times \dfrac{a_1 x_1 + (a_2 x_1)^6}{1 + (a_3 x_1)^6},\quad \theta(x_2) = \bar{\theta}\times \left(1 + \dfrac{(a_4 x_2)^6}{1 + (a_4 x_2)^6}\right).
$$

For the inheritance probability function $p(\vec{x}, \vec{y})$, we assume that the inheritance of $x_1$ acts independently and similarly to the above assumption; however, the inheritance of $x_2$ may depend on the state of the mother cell. Hence, we can write
$$p(\vec{x}, \vec{y}) = \prod_{i=1}^2\dfrac{x_i^{a_i(\vec{y})-1} (1- x_i)^{b_i(\vec{y})-1}}{B(a_i(\vec{y}), b_i(\vec{y}))},$$
where
$$a_1(\vec{y}) = \eta \phi_1(y_1),\quad b_1(\vec{y}) = \eta (1 - \phi_1(y_1)),$$
$$a_2(\vec{y}) = \eta \phi_2(y_1, y_2),\quad b_2(\vec{y}) = \eta (1 - \phi_2(y_1, y_2)).$$
Moreover, the functions $\phi_1$ and $\phi_2$ can depend on the time $t$ in order to specify the changes during development.  

In the simulations, we set $\mu=2.0\times 10^{-4} \mathrm{h}^{-1}, \tau=20 \mathrm{h}, \bar{\theta} = 10^3 \mathrm{cells}, a_1 = 5.8, a_2 = 2.2, a_3=3.75, a_4=2.0, b_1 = 4.0, \eta = 60, \bar{\beta} = 0.12 \mathrm{h}, \kappa_0 = 0.02 \bar{h}^{-1}$, and 
$$\phi_1(y_1) = 0.08 + 1.0 \times \dfrac{(1.65 y_1)^{1.8}}{1 + (1.65 y_1)^{1.8}},$$
$$\phi_2(y_1, y_2) = 0.08 + \left(1 + f(t)\times\dfrac{0.4}{1 + (2.5 y_1)^6}\right) \times \dfrac{(1.56 y_2)^{1.8}}{1 + (1.56 y_2)^{1.8}},$$
where 
$$f(t) = \dfrac{1}{1 + e^{-(t-T_0)/1000}}$$
is a factor that specifies the time-dependent gene regulation. We set $T_0 = 40000$h for the simulations.

\subsection*{\textbf{\textit{Single-cell-based stochastic simulation}}}

In general, the model equation \eqref{eq:2} can be solved numerically through the Euler method and numerical integration. However, when there are two or more epigenetic states, it is expensive to solve the differential integral equation \eqref{eq:s2} numerically because of the high dimensional integration. In this case, we do not solve Eq. \eqref{eq:2} directly, but apply a method of single-cell-based stochastic simulation to model the growth process of a multiple cell system. A multiple cell system is represented as a collection of epigenetic states, and the single-cell-based stochastic simulation tracks the behaviors of each cell according to their own epigenetic states. The sketch of the numerical scheme is given below.
\vspace{0.25cm}

\small{
\begin{minipage}{11cm}
\begin{enumerate}
\item[]
\textbf{Initialize} the time $t=0$, the cell number $Q$, and all cells $\Sigma = \left\{[C_i(\vec{x}_i,a_i)]_{i=1}^{Q}\right\}$. At the initial state, all cells are at the resting phase, and the corresponding age at the proliferating phase (referred to as the age-structured model) is $a_i = 0$. 
\item[]\textbf{for} $t$ from $0$ to $T$ with step $\Delta t$ \textbf{do}
\begin{enumerate}
\item[] \textbf{for} all cells in $\Sigma$ \textbf{do}
\begin{itemize}
\item Calculate the proliferation rate $\beta$, the apoptosis rate $\mu$, and the terminate differentiation rate $\kappa$.
\item Determine the cell fate during the time interval $(t, t+\Delta t)$:  
\begin{itemize}
\item When the cell is at the resting phase, undergo terminal differentiation with a probability $\kappa \Delta t$, or enter the proliferation phase with a probability $\beta \Delta t$. If the cell enters the proliferation phase, set the \textit{age} $a_i=0$.
\item When the cell is at the proliferating phase, if the age $a_i < \tau$, the cell is either removed (through apoptosis) with a probability $\mu \Delta t$ or remains unchanged and $a_i = a_i + \Delta t$; if the age $a_i \geq \tau$, the cell undergoes mitosis and divides into two cells. When mitosis occurs, the epigenetic state of each daughter cell is determined according to the inheritance probability function $p(\vec{x}, \vec{y})$.
\end{itemize}
\end{itemize} 
\item[] \textbf{end for}
\item[] \textbf{Update} the system $\Sigma$ with the cell number, epigenetic states of all surviving cells, and the ages of the proliferating phase cells, and set $t=t + \Delta t$.
\end{enumerate}
\item[]\textbf{end for}
\end{enumerate}
This numerical scheme can be implemented by any object-oriented programming language; we used \texttt{C++} in the current study.    
\end{minipage}}

\section*{Acknowledgements}

This work was supported by the National Natural Science Foundation of China (NSFC 91730301 and 11831015). 


\end{document}